\documentclass[aps,floats,twocolumn,showpacs]{revtex4-1}
\usepackage{amssymb}
\usepackage{amsmath}
\usepackage{graphicx}

 \def\bc{\begin{center}}          \def\ec{\end{center}}
 \allowdisplaybreaks
\begin{document}
 \title{Physics of beam self-modulation in plasma wakefield accelerators}
 \author{K.V.Lotov}
 \affiliation{Novosibirsk State University, 630090, Novosibirsk, Russia}
 \affiliation{Budker Institute of Nuclear Physics SB RAS, 630090, Novosibirsk, Russia}
 \date{\today}
 \begin{abstract}
The self-modulation instability is a key effect that makes possible the usage of nowadays proton beams as drivers for plasma wakefield acceleration. Development of the instability in uniform plasmas and in plasmas with a small density up-step is numerically studied with the focus at nonlinear stages of beam evolution. The step parameters providing the strongest established wakefield are found, and the mechanism of stable bunch train formation is identified.
 \end{abstract}
 \maketitle

\section{Introduction}

Plasma wakefield acceleration driven by proton beams is a concept that may open a path to future high energy lepton colliders \cite{PPCF56-084013}. The interest to this concept is motivated by the ability of plasmas to support strong electric fields and the ability of proton beams to drive these fields over long distances.

Available high energy proton beams consist of rather long (\mbox{$\sim 10$\,cm}) bunches and cannot directly excite plasma waves of the required sub-millimeter length, as the beam frequency spectrum has a negligible component at the plasma frequency. Consequently, the proton bunches must be somehow adapted to the wave. Two methods for that have now been proposed and studied. One method consists in longitudinal compression of the bunch as a whole, which is superior in terms of wakefield excitation \cite{NatPhys9-363,PRST-AB13-041301,SRep4-4171,PRST-AB16-071301}, but is hardly doable within the existing infrastructure \cite{PAC09-4542,PPCF53-014003,AIP1229-510,IPAC10-4395}. Another method involves beam micro-bunching in the plasma due to development of the self-modulation instability (SMI) \cite{EPAC98-806,PPCF53-014003,PRL104-255003}. The latter method is the cornerstone for the first experiment on proton driven plasma wakefield acceleration at CERN named AWAKE \cite{PPCF56-084013,IPAC14-1537,NIMA-740-48,PoP21-123116} and for several supporting experiments with electron beams in other labs \cite{PRL112-045001,NIMA-740-74,AIP1229-467,PoP19-063105,IPAC14-1476}.

The SMI appears due to mutual amplification of the beam radius rippling and the plasma wave \cite{NIMA-410-461}. The linear phase of the wave growth has been studied in several papers and is already well understood. For the case of narrow (compared to the plasma skin depth), low density (compared to the plasma density), constant emittance and constant current beams, analytical expressions for the growth rate and phase velocity of the wave are available \cite{PRL104-255003,PRL107-145003,PRL107-145002,AIP1507-103}, which agree well with computer simulations. The SMI in slightly non-uniform plasmas is analyzed in Refs.\cite{PoP19-010703,PoP20-013102} with the conclusion that the weak non-uniformity does not suppress the self-modulation, though the critical non-uniformity level differs by order of magnitude in theory \cite{PoP19-010703} and simulations \cite{PoP20-013102}. The instability can grow either from an artificially introduced seed perturbation, or from the shot noise that is calculated in Ref.\cite{PRST-AB16-041301}. Various options are possible for the seed perturbation: a short electron bunch \cite{PRST-AB16-041301}, a powerful laser pulse \cite{PoP20-103111}, a sharp cut in the bunch current profile \cite{PoP19-063105}, or a relativistic ionization front co-propagating with the drive bunch \cite{Joshi}. Currently, the co-propagating relativistic front is considered as the mainstream variant, as this solution is chosen for AWAKE experiment.

The seed perturbation becomes a must for beam slicing with the SMI. Otherwise several unstable modes could simultaneously grow and destroy the beam completely \cite{NIMA-410-461,PRL104-255003}. The most dangerous competing mode is the hosing (a non-axisymmetric mode belonging to the same instability family) in which the plasma wave couples with transverse displacements of beam slices. The interplay between hosing and SMI is studied in Refs.\cite{PRE86-026402,PoP20-056704,PRL112-205001} and is by now limited to the analytically tractable case of narrow and constant emittance beams. No detailed numerical studies of mode interaction has been done yet, as this problem is essentially three-dimensional and computationally heavy due to long beams and long interaction distances of interest.

The nonlinear stage of SMI has been studied with numerical simulations only. It was shown that the wakefield amplitude cannot exceed the limit imposed by nonlinear elongation of the wave period \cite{PoP20-083119}, and the designed operation regime of the AWAKE experiment is close to this limit \cite{PoP21-083107}. A finite plasma radius or weak radial non-uniformity of the plasma have a small effect on the wave growth, if the radial scale of density variation is much greater than the plasma skin depth \cite{PoP21-056703,PRL112-194801}. As self-modulating beams are many plasma wavelengths long, motion of plasma ions could come into play and hamper wave excitation, if plasma ions are light \cite{PRL109-145005,PoP21-056705}.

Once the self-modulation is completed, that is, the beam is split into well separated micro-bunches, the beam evolution continues towards some strongly destroyed state \cite{PoP18-024501}. As a result, the wakefield amplitude as a function of the propagation distance drops down soon after reaching the maximum \cite{PoP18-103101}. As a cure for beam destruction, the small step in plasma density was proposed in Ref.\cite{PoP18-024501}. With the density step, the wakefield amplitude remains at some high level for a long distance \cite{PoP18-103101}. A proper longitudinal non-uniformity of the plasma density is thus a necessary element of a future plasma wakefield accelerator if the latter is based on self-modulation of long proton bunches.

Optimum locations and magnitudes of density steps in Refs.\cite{PoP18-024501,PoP18-103101} were found by means of a straightforward numerical optimization without deep studies of how the step works and why its optimum is there. In this paper, we clarify mechanisms of beam destruction in the uniform plasma and the role played by the density step. Wherever possible, we compare two cases (the uniform plasma and the optimum density step) to emphasize the effect of the step on beam evolution. In Sec.\ref{s2} we define the problem under study. In Sec.\ref{s3} we describe macroscopic properties of beam evolution observed in simulations. Then we focus on the motion of beam particles in Sec.\ref{s4} and analyze contributions of individual micro-bunches in Sec.\ref{s5}. Main findings are summarized in Sec.\ref{s6}.

\section{Statement of the problem} \label{s2}

We use units of measure that are traditional for plasma wakefield studies: speed of light $c$ for velocities, electron mass $m$ for masses, initial plasma density $n_0$ for densities, inverse plasma frequency $\omega_p^{-1}$ for times, plasma skin depth $c/\omega_p$ for distances, and wavebreaking field $E_0 = mc\omega_p/e$ for fields. The SMI perturbs the beam axisymmetrically, so cylindrical coordinates $(r, \varphi, z)$ and the co-moving coordinate $\xi = z - ct$ are convenient.

As we aim at the detailed study of the nonlinear instability stage, we intentionally exclude all irrelevant effects by the choice of beam and plasma parameters. In particular, we consider the radially uniform cold plasma with immobile ions and the mono-energetic positron beam with the density profile
\begin{equation}\label{e1}
    n_b = \begin{cases}
        n_{b0} e^{-r^2/(2 \sigma_r^2)},\quad & \xi<0, \\
        0, & \xi\geq 0,
    \end{cases}
\end{equation}
where $n_{b0} = 4 \times 10^{-3}$ and $\sigma_r = 0.5$. The hard leading edge seeds the instability in the same way as in most discussed experiments. Unless stated otherwise, we simulate about 25 wave periods, so the plasma response to the beam is surely linear for the chosen beam density. The positive charge of the beam makes discussions of potential wells more intuitive. The beam radius is chosen as a compromise between the analytically tractable case $\sigma_r \ll 1$ and the design value of AWAKE experiment $\sigma_r = 1$. The beam relativistic factor $\gamma_b = 1000$ is high enough to separate the betatron frequency and the plasma frequency. Using light particles (positrons) greatly speeds up simulations as compared to the proton beams. The beam angular spread $\delta \alpha = 2 \times 10^{-4}$ is low enough to avoid the emittance-driven beam divergence that would otherwise complicate the self-modulation \cite{PoP18-103101}.

To ease comparison with earlier theoretical results, we also introduce the inverse betatron frequency
\begin{equation}\label{e2}
    \tau_b = \sqrt{2 \gamma_b / n_{b0}} \approx 707.1,
\end{equation}
which is the unit of time, e.g., in Refs.\,\cite{PRL104-255003,PRL107-145003}.

\begin{figure}[tbh]
\includegraphics[width=218bp]{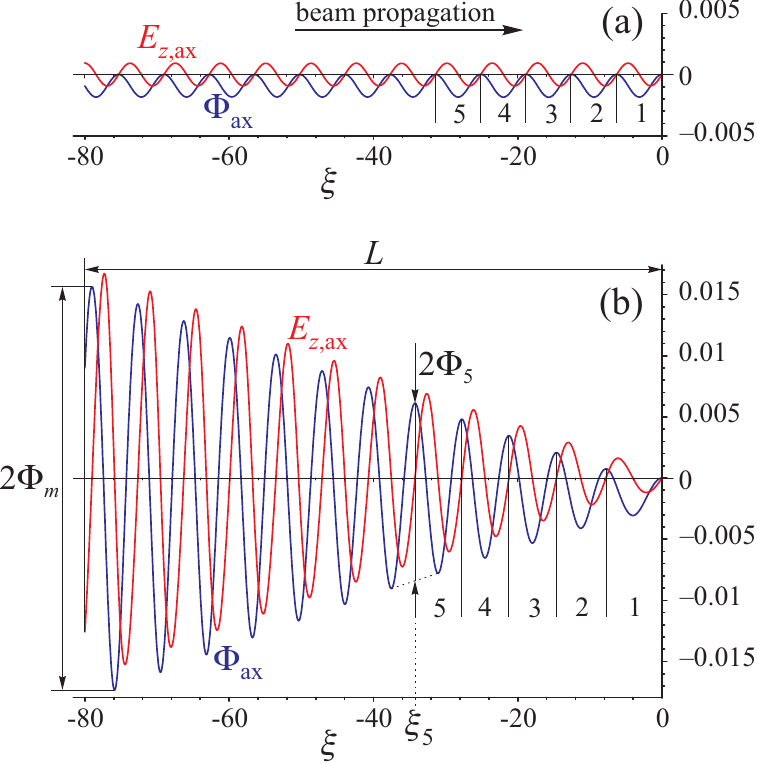}
\caption{On-axis electric field $E_{z,\text{ax}}$ and wakefield potential $\Phi_\text{ax}$ of the plasma wave at two propagation distances: the beginning of the interaction at $z=0$ (a) and a place of the almost maximum wakefield in the uniform plasma at $z=1000\approx \sqrt{2}\tau_b$ (b). Regions of first five bunches are indicated by vertical lines and numbers. Definitions for $\xi_j$ and $\Phi_j$ are illustrated by the example of the fifth bunch.}\label{f1-pattern}
\end{figure}
Since the plasma response is linear and axisymmetric, we simulate the SMI with the axisymmetric quasi-static 2d3v hybrid code LCODE \cite{PoP5-785,LCODE}, in which the beam is modeled with macro-particles, and plasma is treated as the electron fluid. In some cases the result is cross-checked with the much slower fully kinetic version of LCODE \cite{PRST-AB6-061301,IPAC13-1238}.

\begin{figure}[tbh]
\includegraphics[width=217bp]{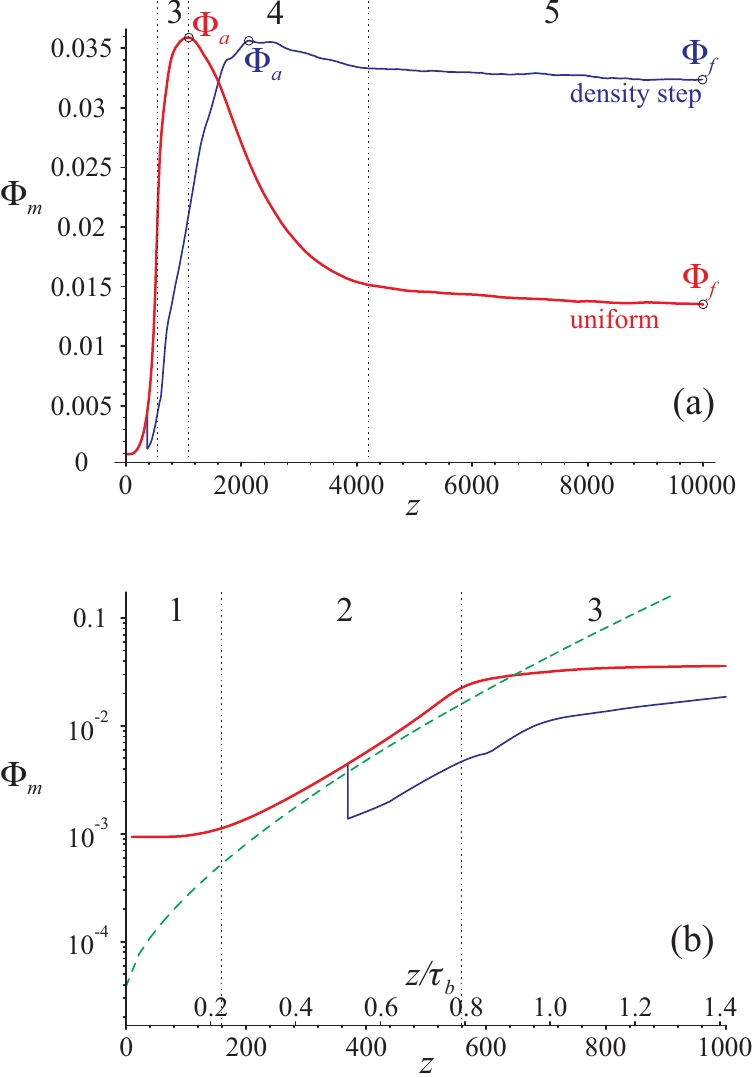}
\caption{The maximum wakefield amplitude $\Phi_m$ as a function of propagation distance $z$ for the uniform plasma (red lines) and the plasma with the optimum density step (blue lines) in normal (a) and enlarged semi-logarithmic (b) scales. The green dash line in (b) shows the theoretically predicted growth. The numbers and vertical lines indicate stages of beam evolution in the uniform plasma. Circles in (a) show points of absolute maximum $\Phi_a$ and established amplitude $\Phi_f$.  }\label{f2-Phi_m}
\end{figure}

Typical patterns of the plasma wake are shown in Fig.\,\ref{f1-pattern}, where the subscripts ``ax'' denote on-axis values of the quantities. Along with the longitudinal electric field $E_z$ on the axis, we also show the on-axis value of the wakefield potential
\begin{equation}\label{e3}
    \Phi (r, \xi) = \int_\xi^\infty E_z (r, \xi') \, d \xi'.
\end{equation}
The force acting on an axially moving ultrarelativistic particle is just ($-\nabla \Phi$). Regions of negative $\Phi$ are thus the potential wells that can confine beam particles.

It is natural to consider parts of the beam which lie between adjacent maxima of $\Phi_\text{ax} (\xi)$ as ``bunches''. This definition is convenient in that the bunches can be identified even at early stages of their formation. We define the bunch coordinate $\xi_j$ as the location of local potential maximum at the ``end'' of the bunch and denote the wakefield amplitude after passage of the bunch by $\Phi_j$, where $j$ is the bunch number. These definitions are illustrated in Fig.\,\ref{f1-pattern}(b). The chosen definition of $\Phi_j$ (more complex one than simply a value) is necessary for excluding the slowly varying component of the wakefield which appears due to the average beam current. We also introduce the maximum wakefield amplitude $\Phi_m (z,L)$ that we consider as a function of propagation distance $z$ and beam length $L$. In what follows we put $L=160$ unless a dependence on $L$ is studied.

\section{Phenomenology of self-modulation} \label{s3}

In this section, we describe general SMI features and introduce the terminology necessary for further studies.

First we consider the maximum wakefield amplitude $\Phi_m$ in relation to the propagation distance $z$ (or the propagation time, which is the same) for the uniform plasma and the plasma with the optimum density step (Fig.\,\ref{f2-Phi_m}). Here the optimum step  means a steep increase of the plasma density by 8.5\% at $z=360$. This combination of numbers maximizes the established wakefield amplitude that we measure at $z=10000$. Quantitative characteristics of the presented graphs are beam parameter dependent, so we focus on the qualitative behavior that is typical for the SMI.

In Fig.\,\ref{f2-Phi_m} we can distinguish five stages of beam evolution in the uniform plasma. At the 1st stage, the wakefield structure changes from that of the seed perturbation to that corresponding to fastest instability growth. Then there is the 2nd stage characterized by the nearly exponential growth of the wakefield amplitude. This growth closely fits the theoretically predicted rate \cite{PRL107-145003}
\begin{equation}\label{e4}
    \Phi_m \propto \exp \left( \frac{3 \sqrt{3}}{4} L^{1/3} \left( \frac{z}{\tau_b} \right)^{2/3} \right),
\end{equation}
which is shown in Fig.\,\ref{f2-Phi_m}(b) by the dash line. The 2nd stage gives place to the 3rd stage (non-exponential growth) well before the maximum amplitude is reached. The 4th stage is that of the fast field decrease, and the 5th stage is characterized by the almost constant wakefield. With the density step, these stages are also observed, though the non-exponential growth is longer, and the established wakefield amplitude is higher.

\begin{figure}[bt]
\includegraphics[width=192bp]{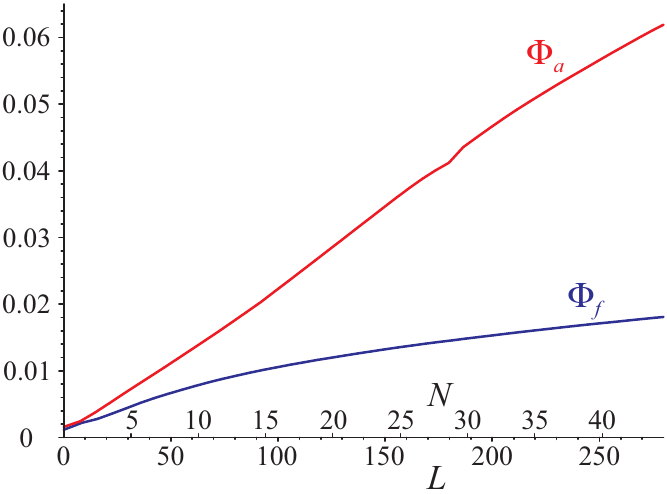}
\caption{Growth of the wakefield amplitude with the beam length $L$ and the number of passed bunches $N$ at the moment of maximum wakefield ($\Phi_a$) and in the established equilibrium state ($\Phi_f$).}\label{f3-Phi(L)}
\end{figure}

The graphs $\Phi_m(z)$ in Fig.\,\ref{f2-Phi_m}(a) contain two significant points each: the absolute maximum of the amplitude $\Phi_a$ and the established amplitude $\Phi_f$. In the uniform plasma, both amplitudes monotonically grow with the beam length $L$ (Fig.\,\ref{f3-Phi(L)}) and, therefore, show how individual bunches participate in driving the wave. At the point of maximum field ($\Phi_a$), all available bunches work equally. At the established state ($\Phi_f$), contributions of bunches decrease with the bunch number $N$. In other words, the longer the beam, the smaller the ratio of the established field to the peak field. There is no sense in plotting similar graphs for stepped density profiles, as the choice of the optimum case depends on the number of bunches.

\begin{figure}[tbh]
\includegraphics[width=210bp]{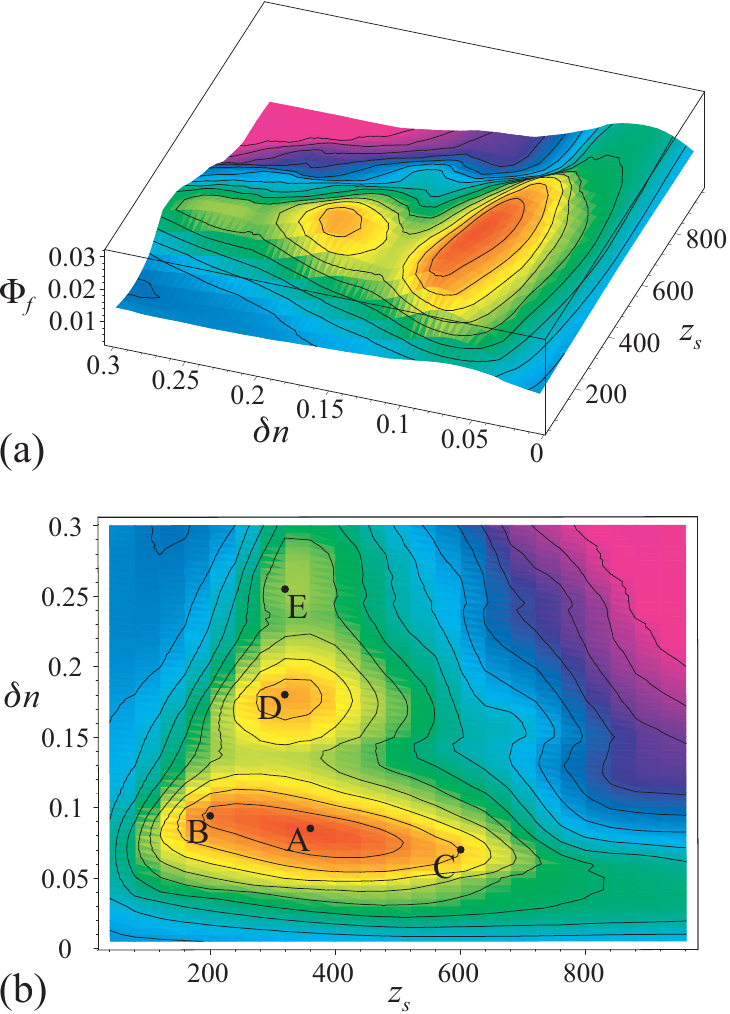}
\caption{The map of step efficiencies $\Phi_f (\delta n, z)$ in two projections.}\label{f4-map}
\end{figure}
\begin{figure}[tbh]
\includegraphics[width=201bp]{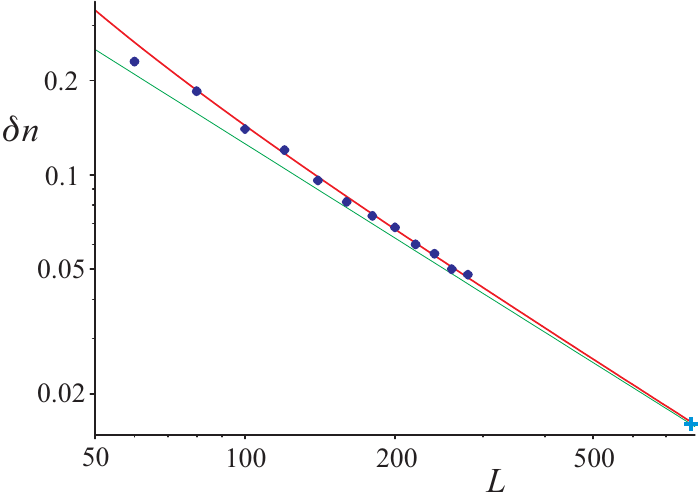}
\caption{The magnitude of the optimum density step $\delta n$ as a function of the beam length $L$. Points are obtained from simulations, lines are approximations. }\label{f5-dn(L)}
\end{figure}

The purpose of the density step is to make the established wakefield amplitude as high as possible. We can vary location $z_s$, magnitude $\delta n$, and steepness of the step in search of maximum $\Phi_f$. For the sharp density step and $L=160$, the function $\Phi_f (\delta n, z)$ is shown in Fig.\,\ref{f4-map}. We see that the area of optimum parameters extends over a wide interval in $z_s$ covering the whole 2nd (exponential) stage of beam evolution [compare with Fig.\,\ref{f2-Phi_m}(b)]. The wakefield amplitude reached before the step varies over an order of magnitude in this interval. The required magnitude of the step, however, is well defined for each step location. There are also several higher order maxima at multiples of the optimum step magnitude.

\begin{figure}[tbh]
\includegraphics[width=201bp]{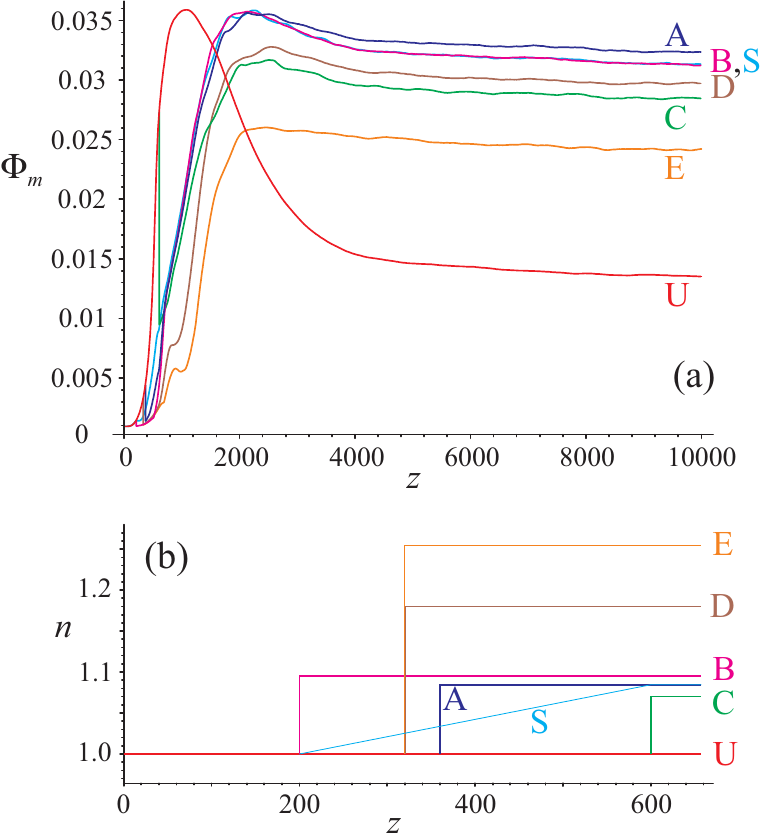}
\caption{The maximum wakefield amplitude $\Phi_m$ as a function of propagation distance $z$ (a) for selected longitudinal plasma profiles (b).}\label{f6-steps}
\end{figure}

The optimum magnitude of the plasma density step $\delta n$ depends on the beam length $L$ or, equivalently, the number of bunches $N \approx L/(2 \pi)$ (Fig.\,\ref{f5-dn(L)}). There is a simple engineering formula for this dependence
\begin{equation}\label{e5}
    \delta n \approx \frac{4 \pi}{L - 4 \pi} \approx \frac{2}{N-2}
\end{equation}
shown by the thick red line in Fig.\,\ref{f5-dn(L)}. The inverse dependence on $L$ is expectable, as it provides the same average shift of the wave with respect to beam for beams of different lengths. The numerical factor in this dependence means that the optimum step makes the beam exactly one plasma period longer, if measured in local plasma wavelengths. The correction subtrahend in the denominator indicates that the first two bunches play a special role in the self-modulation. Without this subtrahend, the approximation $\delta n = 2/N$ deviates from simulations at small $L$ (thin green curve in Fig.\,\ref{f5-dn(L)}). Earlier simulations of low-emittance beams also fit the formula (\ref{e5}): the rightmost (cruciform) point is that for LHC beam \cite{PoP18-103101} with $\sigma_{z,0}$ taken as $L$.

Figure~\ref{f6-steps}(a) shows the time evolution of the maximum wakefield amplitude $\Phi_m$ for several typical steps. These steps are drawn in scale in Fig.\,\ref{f6-steps}(b) and indicated in Fig.\,\ref{f4-map}(b) with the same characters. We see that the maximum field behaves similarly at different points of the ``optimum plateau'' and at higher order maxima. The smooth density increase (line ``S'') produces almost the same result as the sharp step located at the beginning of the increase (line ``B''). We therefore concentrate on steep steps, as a reasonably short length of the density transition has a little effect on the resulting field.

\begin{figure*}[p]
\includegraphics[width=281bp]{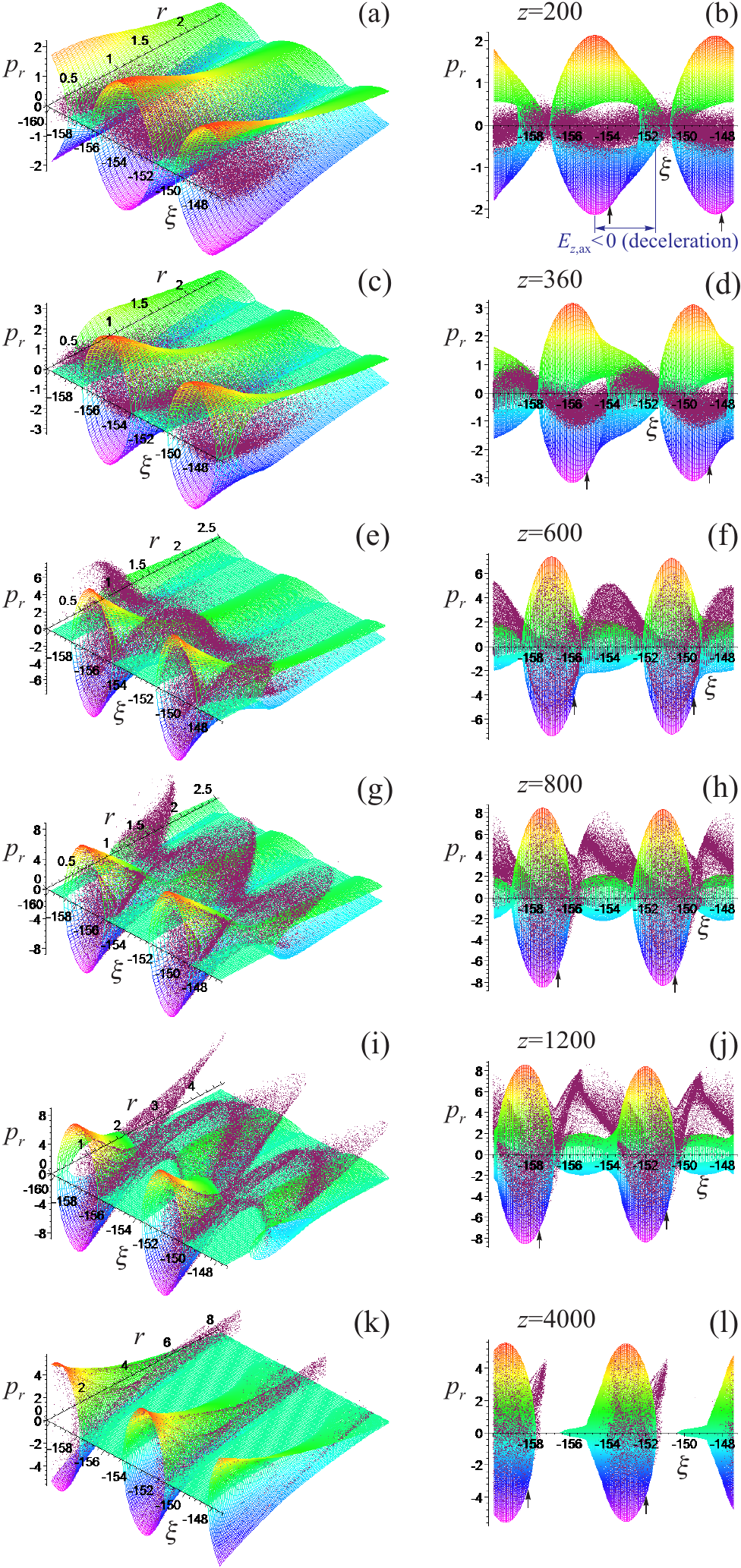}
\caption{Beam particles (points) and the separatrix (wireframe surface) in $(r, \xi, p_r)$ space at different times for the uniform plasma case. In right fragments, small arrows show locations of the maximum on-axis beam density.}\label{f7-uniform}
\end{figure*}

\section{Motion of beam particles} \label{s4}

The bunches are created and then partly destroyed because of the radial motion of particles in potential wells of the wakefield. A slow longitudinal motion of potential wells with respect to particles is also important. We therefore look at the motion of beam particles in the space $(r, \xi, p_r)$, where $p_r$ is the radial momentum. The azimuthal angle $\varphi$ is not important due to the axial symmetry, the conserved azimuthal momentum $p_\varphi$ is negligible at low emittances, and the longitudinal momentum $p_z$ is almost constant at the considered time intervals. The potential wells in this space can be conveniently visualized by the separatrix, which is the surface separating particles of positive and negative energy of the transverse motion. The separatrix is defined as
\begin{equation}\label{e6}
    p_r = \pm \sqrt{-2 \gamma_b \Phi (r, \xi)} \quad \text{at} \quad \Phi (r, \xi) \leq 0
\end{equation}
and usually has the shape of a cone with negative-energy particles inside. We consider the interval $-160<\xi<-147.8$ and display the resulting three-dimensional portraits of the beam and the separatrix in two projections at several time slices.

The case of the uniform plasma is shown in Fig.\,\ref{f7-uniform}. At 1st and 2nd stages of beam evolution [fragments (a)-(d)], the transverse motion of particles is field dominated. The separatrix width scales as the square root of the wave amplitude, while the particle transverse momentum is directly proportional to the amplitude, so the separatrix is much wider than the beam footprint at low amplitudes [Fig.\,\ref{f7-uniform}(b,d)]. This means the transverse momentum of a particle is much smaller than that needed for escaping the potential well or than the momentum a particle can gain falling into the potential well. Since the wave amplitude quickly grows during these stages, we can also neglect the momentum imparted to the particle at earlier times when the potential well had a different shape or location. Simply stated, at any time we can analyze the particle motion as if the particle recently appeared in the potential well with zero momentum.

\begin{figure}[tbh]
\includegraphics[width=205bp]{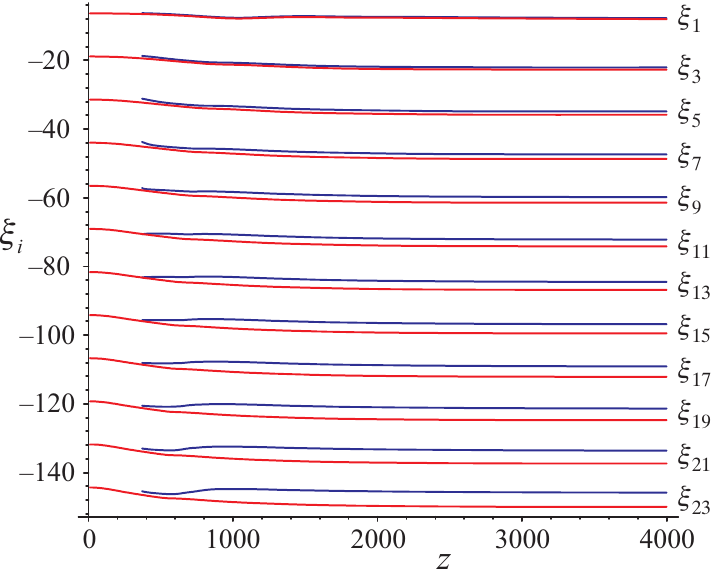}
\caption{Bunch coordinates $\xi_j$ as functions of the propagation distance $z$ for the uniform plasma (red) and the optimum step (blue). Every second bunch is shown. }\label{f8-phases}
\end{figure}

Regions of the decelerating field in the plasma wave are shifted a quarter-period with respect to potential wells. In the right column of Fig.\,\ref{f7-uniform}, this is the right (forward) shift [see Fig.\,\ref{f7-uniform}(b)]. The beam drives the wave only if there are more decelerated particles than accelerated ones; otherwise the wave would damp as the bunches pass by. Consequently, the potential wells must be populated asymmetrically with beam particles. The asymmetry is achieved by the backward motion of potential wells with respect to the beam (Fig.\,\ref{f7-uniform}, right column). The mechanism of the motion we will discuss later. With this motion, the focusing force first tightens the beam in some place, and then the well falls back leaving the densest part of the beam (marked by small arrows in Fig.\,\ref{f7-uniform}) in the decelerating field. This type of wave-particles interaction turns out to be the most efficient way of wave excitation, so it dominates over other possible perturbations and determines the overall beam evolution. The backward motion of the wave was characterized in Refs.\cite{PRL107-145003,PRL107-145002} in terms of the phase velocity.

As the potential well moves with respect to the beam, it ingests beam particles at one side and releases at the other. This process continues while the transverse particle motion is negligibly slow, but eventually comes to the end due to the increased particle momentum or radial displacement of particles [Fig.\,\ref{f7-uniform}(e)-(h)]. Once this happens, the exponential growth (2nd stage) changes to the non-exponential growth (3rd stage). The reduction of particle inflow to the potential wells favors the asymmetry of well population, so the wave amplitude continues to grow. The growth rate, however, is smaller, as the unbalanced escape of particles from the well results in progressive well depopulation.

The wakefield amplitude reaches its maximum when the decelerating phase of the wave approaches the particle void area [Fig.\,\ref{f7-uniform}(i,j)]. At this moment, a considerable number of particles is in the strong decelerating field, but some of them are outside the separatrix. These particles gradually move aside from the axis causing the wakefield reduction during the 4th stage. This is a positive feedback effect, as the reduced wakefield results in additional losses of particles from the well. Another reason for the extra particle loss is the motion of the potential well that continues during the 4th stage [Fig.\,\ref{f7-uniform}(j,l)]. At the stage of established wakefield (the 5th stage), potential wells and particle distributions inside the wells change a little compared to Fig.\,\ref{f7-uniform}(k,l).

The motion of potential wells with respect to the co-moving frame is quantitatively characterized by Fig.\,\ref{f8-phases}. The motion with respect to beam particles is almost the same, as the average particle  $z$-velocity (determined by the transverse momentum) differs from the light velocity by the term of the order of $p_r^2/(2\gamma_b^2) \lesssim 10^{-5}$. In the uniform plasma, the motion stops at approximately the middle of the 4th (field decrease) stage. The integral shift is expectedly larger for tail bunches.

\begin{figure*}[p]
\includegraphics[width=281bp]{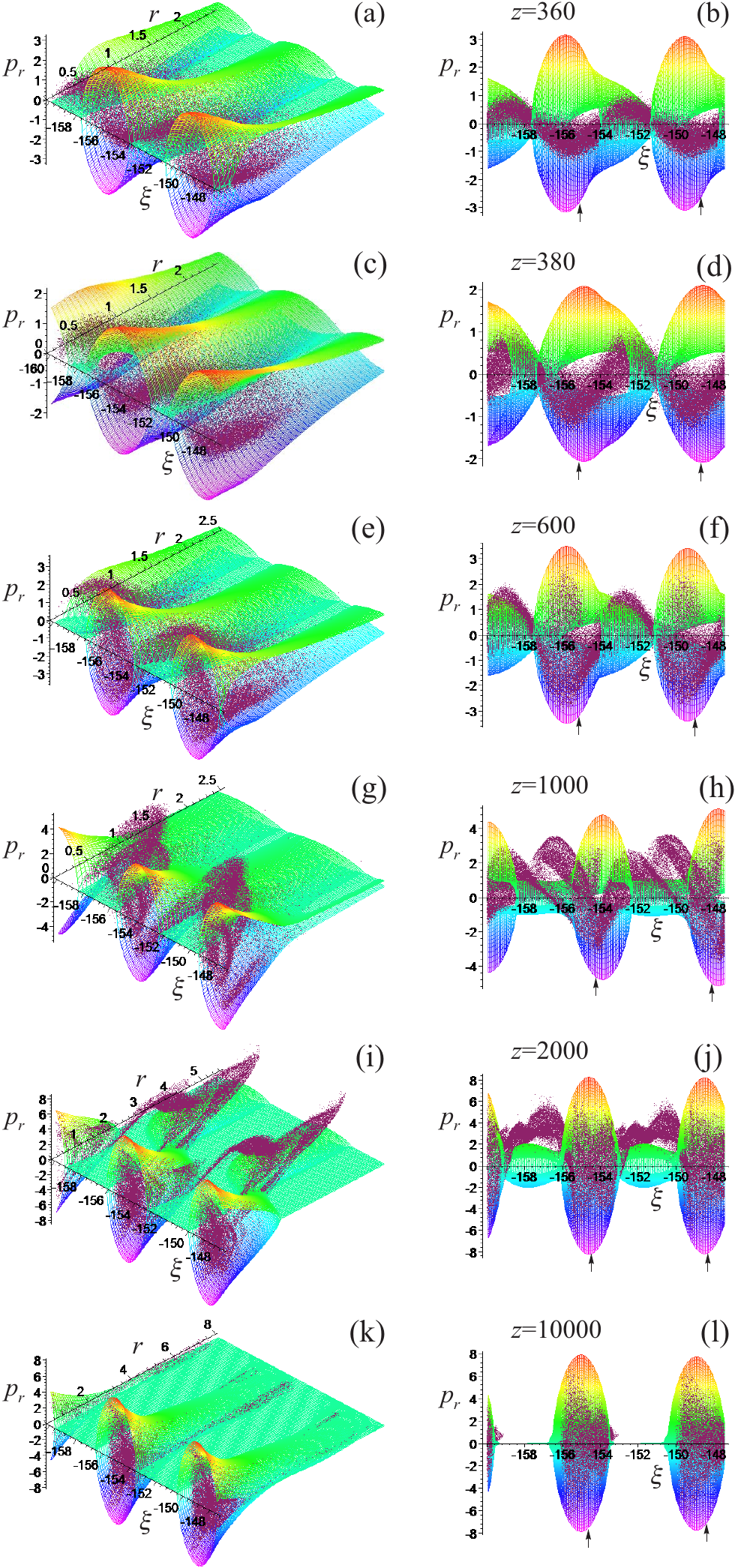}
\caption{Beam particles (points) and the separatrix (wireframe surface) in $(r, \xi, p_r)$ space at different times for the optimum density step. In right fragments, small arrows show locations of the maximum on-axis beam density.}\label{f9-stepped}
\end{figure*}
\begin{figure*}[tbh]
\includegraphics[width=426bp]{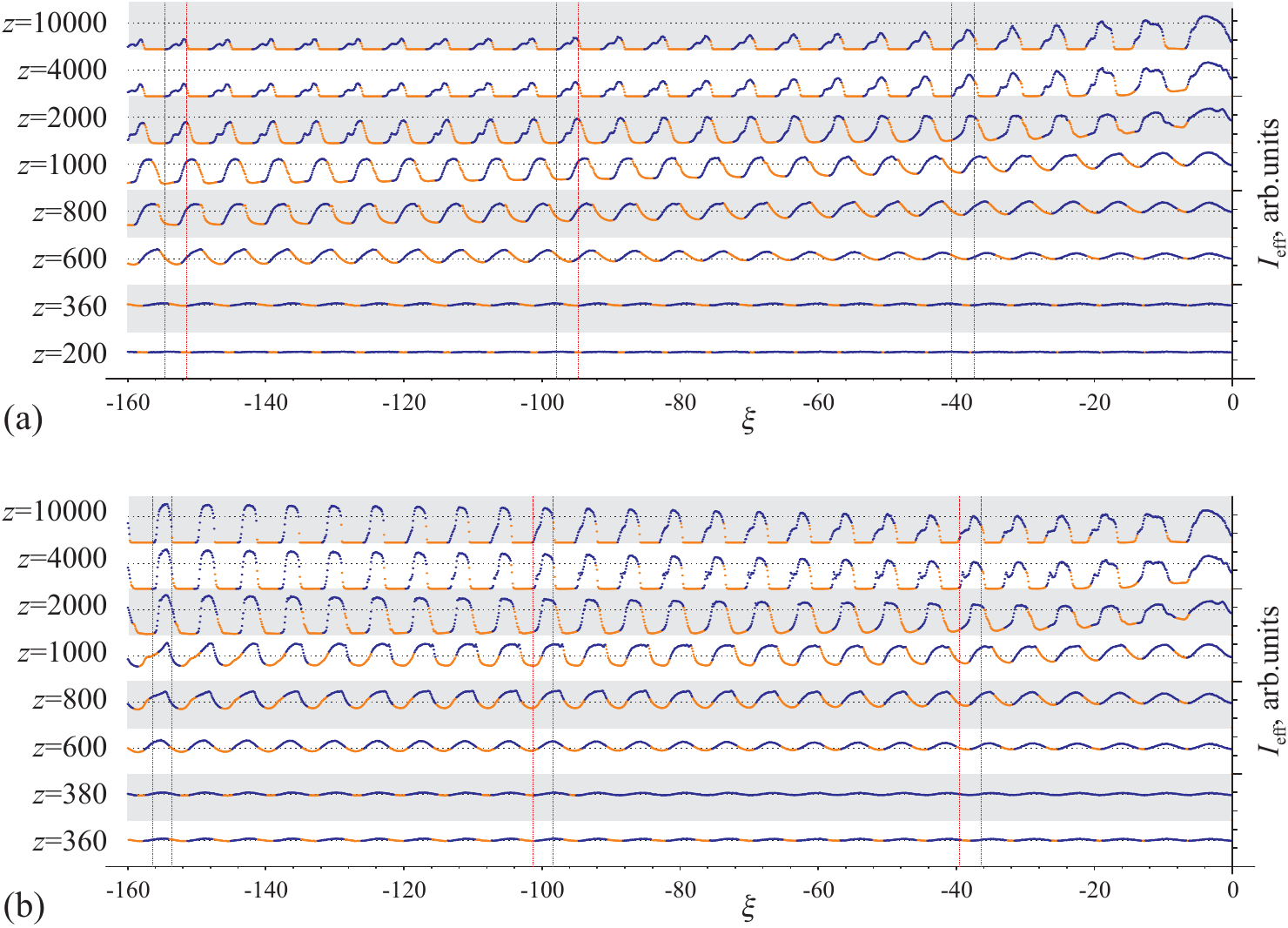}
\caption{The effective beam current $I_\text{eff} (\xi)$ at several propagation distances $z$ in the uniform (a) and stepped-up (b) plasmas. The zero current is at the bottom of each drawing zone. Blue parts of the curves correspond to focusing phases of the wakefield ($\Phi_\text{ax}<0$), orange parts to defocusing ($\Phi_\text{ax}>0$). Horizontal dotted lines show the effective current of the initial beam. Red vertical lines help to relate locations of bunches at different times with their final locations.}\label{f10-current}
\end{figure*}

Beam and separatrix portraits for the case of the optimum density step (profile ``A'' in Fig.\,\ref{f6-steps}) are shown in Fig.\,\ref{f9-stepped}. Snapshots (a) and (b) are taken right before the density step and are the same as in Fig.\,\ref{f7-uniform}(c,d), fragments (c) and (d) are taken soon after. The step makes the plasma wavelength shorter, but does not change the number of wakefield periods (Fig.\,\ref{f8-phases}). The partially bunched (or rather, rippled) beam thus drives the wave with the period that is substantially longer than the local plasma wave period. The increased density acts as a ``spring'' resisting further elongation of the wakefield period. If the ``spring pressure'' is properly adjusted by the step magnitude, then the wake period does not change, and potential wells do not move with respect to the beam (Fig.\,\ref{f8-phases}, right column of Fig.\,\ref{f9-stepped}). The strength of the ``spring'', however, depends on the wave amplitude: the higher the stronger. This is because the own wakefield of a particle bunch has a little effect on the high amplitude wave, but easily modifies a weak wave. We therefore observe some backward shift of the wave in the time interval of the low-amplitude wave ($z \lesssim 800$ in Fig.\,\ref{f8-phases}).

There is, however, the question of how to provide the asymmetry of well population necessary for efficient driving the wave. As we see from Fig.\,\ref{f9-stepped}, the backward shift of the wave at $z \sim 600$ is responsible for that. The shifted potential well imparts a large transverse momentum to some beam particles [Fig.\,\ref{f9-stepped}(e,f)] that fall outside the separatrix when the well shifts forward [Fig.\,\ref{f9-stepped}(g-j)] and create a gap in the established well population [Fig.\,\ref{f9-stepped}(k,l)].

Evolution and phasing of the bunch train as a whole can be examined with the help of the effective current
\begin{equation}\label{e7}
    I_\text{eff} (\xi) = \int_0^\infty n_b (r, \xi) e^{-r} \, 2 \pi r \, dr
\end{equation}
that qualitatively accounts for reduced contribution of defocused particles to the wakefield. Strictly speaking, the contribution of an off-axis particle to the on-axis electric field $E_z$ is proportional to the modified Bessel function of the second kind $K_0 (r)$ \cite{PAcc22-81,PRST-AB16-041301}. However, the exact formula is impractical when applied to simulations output because of the noise generated by near-axis particles, so we use (\ref{e7}). It is of particular interest to correlate the effective current and the location of focusing and defocusing regions.

The map of effective beam currents for the uniform plasma [Fig.\,\ref{f10-current}(a)] reveals that all bunches in the beam, except the first one, behave similarly: leading edges of the bunches are in defocusing phases, trailing edges are focused, and the bunches move backward in $\xi$ until their trailing edges run into zero $I_\text{eff}$ regions. After that, the defocusing phases ``eat up'' large parts of the bunches, the larger the further the bunch is from the beam head.

In the stepped-up plasma [Fig.\,\ref{f10-current}(b)], we observe the transition between two typical cases as we go from head to tail of the beam. For the head bunches, the considered density step is smaller than the optimum one, so the bunches evolve almost as in the uniform plasma: the final bunch shapes at $\xi \gtrsim -40$ in fragments (a) and (b) of Fig.\,\ref{f10-current} are similar, and the bunches move backward at the nearly the same speed. For the tail bunches ($\xi \sim -160$), the backward motion is suppressed by the reduced plasma wavelength, but the bunches in the final state fill potential wells almost symmetrically and drive the wave inefficiently. In the middle ($\xi \sim -100$), the final bunch shape looks optimal, as decelerated particles survive, while accelerating phases of the wave are depleted.

As we see, the density step instantly switches the beam-plasma system from the fastest growing mode to the state characterized by a slower wave growth and a favorable final beam state. This is possible owing to the ability of the grown SMI mode to suppress development of other modes. The same feature enables suppression of the hosing in the case of a seeded SMI. Otherwise, growth of the detuned mode would be hampered by the fastest mode at the new plasma density.

The optimum switching occurs at the linear stage of the wave growth at which the ratio of wave amplitudes before and after the step does not depend on the amplitudes themselves. This is the reason why the interval of possible step locations is as long as the whole second stage of beam evolution.

\section{Contributions of separate bunches} \label{s5}

In the previous section, we looked at the process from the beam's standpoint and mostly examined how the beam is changed by the wakefield. Here we take the plasma wave as a reference and consider the complementary part of the process, the wakefield excitation by the bunched beam.

\begin{figure}[tbh]
\includegraphics[width=218bp]{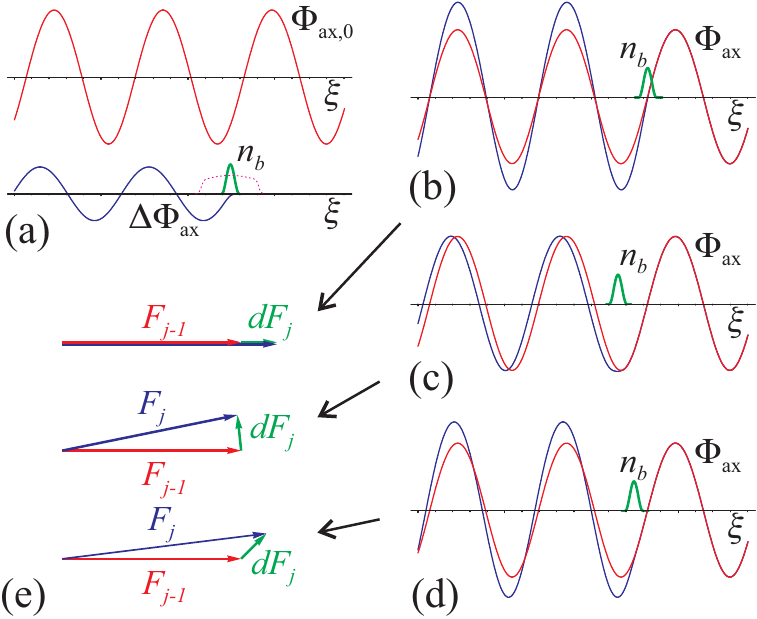}
\caption{Illustration of wakefield modification by a short bunch. Wakefield of the bunch is: (a) drawn separately, (b) in phase with the master wave, (c) approximately $\pi/2$ delayed, and (d) $\pi/4$ delayed. (e) Composition of complex amplitudes for the three cases. }\label{f11-tutor}
\end{figure}

We start with reminding few important consequences of the linear wakefield theory \cite{PF30-252}. A short particle bunch traveling in the plasma wave modifies this wave by its own wakefield (Fig.\,\ref{f11-tutor}). The result of modification depends on which phase of the wave the bunch is located at. If the bunch is at the maximum decelerating gradient, its wakefield is in phase with the master wave, and the bunch simply amplifies the  wave [Fig.\,\ref{f11-tutor}(b)]. If the bunch is located at the bottom of the potential well, the wave amplitude remains unchanged, but the phase is advanced [Fig.\,\ref{f11-tutor}(c)]. At intermediate locations, the bunch simultaneously amplifies the wave and moves it backward [Fig.\,\ref{f11-tutor}(d)]. Note that same conclusions are true for elongated symmetric bunches like that shown by the dash line in Fig.\,\ref{f11-tutor}(a).

The above considerations explain why the wave motion with respect to the beam continues after reaching the maximum wave amplitude in the uniform plasma. At the maximum amplitude, front parts of the bunches are in the defocusing field [Fig.\,\ref{f7-uniform}(i,j), Fig.\,\ref{f10-current}(a)]. These parts must disappear and thereby shift bunch centroids backward, thus shifting the wave backward.

The wave modification by individual bunches can be conveniently characterized by the complex amplitude \begin{equation}\label{e8}
    F_j = \Phi_j \exp (-i \xi_j \sqrt{n}),
\end{equation}
where $j$ is the bunch number, and $n$ is the local plasma density that affects the local plasma wavelength. Each bunch makes the increment $\Delta F_j = F_j - F_{j-1}$ to the complex amplitude. It is more informative, however, to look at the quantity
\begin{equation}\label{e9}
    dF_j = \Delta F_j \exp (i \xi_{j-1} \sqrt{n})
\end{equation}
that characterizes the bunch contribution in relation to the wave excited by previous bunches [Fig.\,\ref{f11-tutor}(e)].

\begin{figure}[b]
\includegraphics[width=172bp]{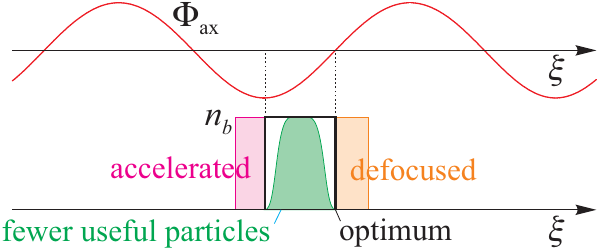}
\caption{Illustration of the optimum bunch positioning. }\label{f12-optimum}
\end{figure}
\begin{figure*}[tbh]
\includegraphics[width=438bp]{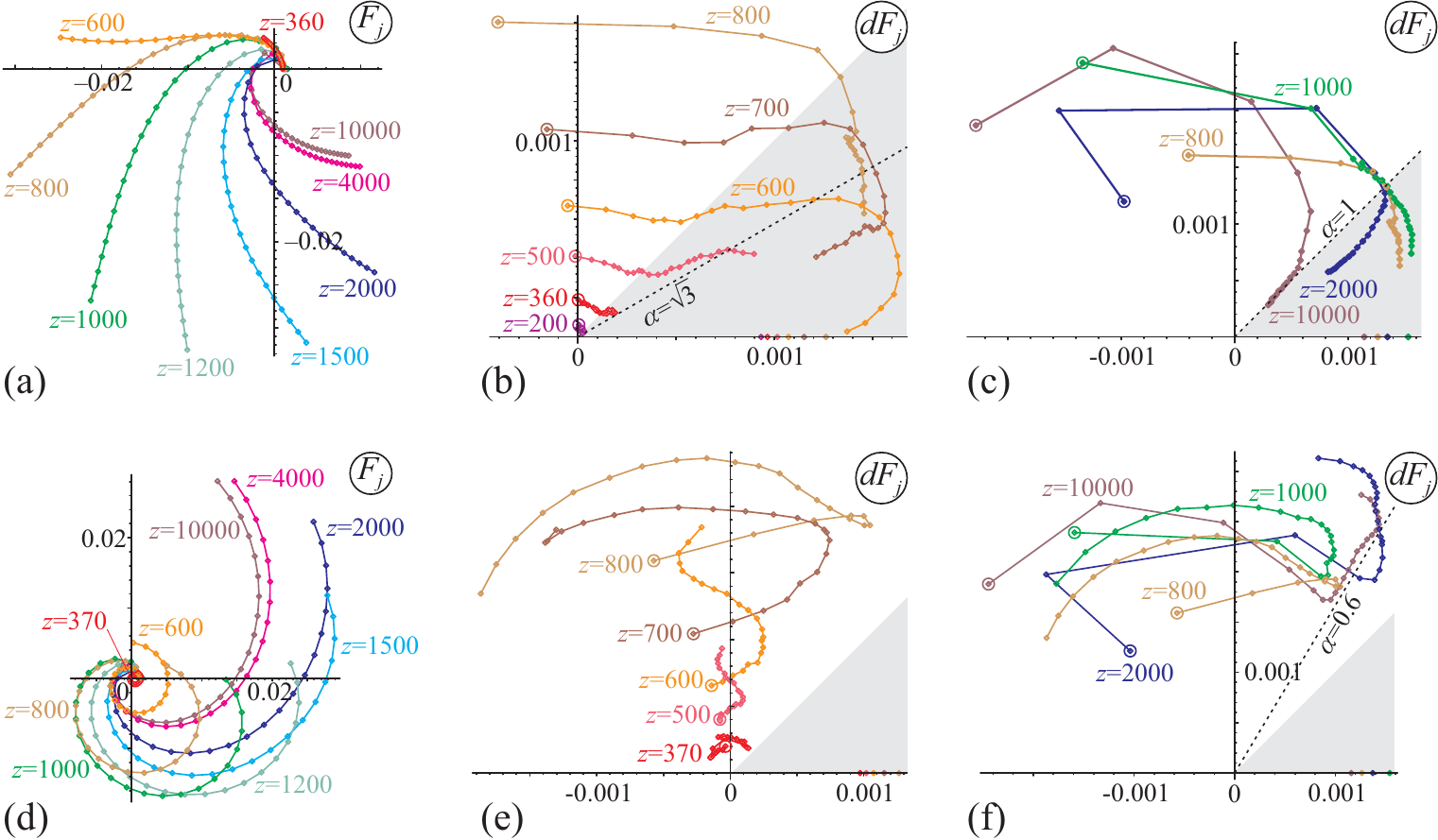}
\caption{The complex amplitude $F_j$ (a,d) and the increments $dF_j$ to the complex amplitude (b,c,e,f) at different propagation distances for uniform (a,b,c) and stepped-up (d,e,f) plasmas. Points of the same color correspond to different bunch numbers $j$, lines between them are drawn for better visibility. In fragments (b,c,e,f), small circles mark contributions of the second bunch, points on the abscissa axis are contributions of the first bunch, gray shadings show areas of defocusing risk, and dashed lines are theoretical approximations explained in the text.}\label{f13-spiral}
\end{figure*}

If the bunch is a fragment of some constant-current long beam, then its contribution to the existing wakefield is strongest if the bunch completely fills the decelerating and focusing phase of the wave (Fig.\,\ref{f12-optimum}). A shorter or partially destroyed bunch is obviously less efficient. A longer bunch contains either accelerated particles that reduce the wave or defocused particles that make the bunch unstable. The optimum angle between $dF_j$ and $F_{j-1}$ is thus $\pi/4$, as in Fig.\,\ref{f11-tutor}(d).

Growth of complex amplitudes at different propagation distances is shown in Fig.\,\ref{f13-spiral}(a,d) for the two studied cases. With the density step, the spirals are expectedly tighter, since the ratio of the bunch-to-bunch distance to the plasma period is higher in this case. Other differences are seen only at the graphs of individual bunch contributions. In the uniform plasma, the five stages of beam evolution correspond to qualitatively different behavior of $dF_j$. At the first stage ($z \lesssim 200$), the bunches appear where the potential wells of the seed wakefield are, and their contributions are purely imaginary: no amplitude increase, only the phase shift [Fig.\,\ref{f13-spiral}(b)]. The exponential growth stage ($200 \lesssim z \lesssim 600$) is characterized by the nearly constant phase shift (imaginary part) and the amplitude increment (real part of $dF_j$) that grows along the beam. The transition from the exponential to the non-exponential growths ($z \sim 600$) occurs when contributions of tail bunches approach that of a fully bunched beam ($|dF_j|\sim 0.002$ in our case). As soon as the bunch contribution cannot grow further in absolute value, it moves down to the area of defocusing danger [gray areas in Fig.\,\ref{f13-spiral}(b,c)]. In other words, the bunch advances the wave phase less and therefore increases the local phase velocity of the wave. Staying in the gray area is accompanied with loss of particles, so the bunch contribution reduces in the absolute value at $z \gtrsim 1000$ [Fig.\,\ref{f13-spiral}(c)] until $dF_j$ exits the defocusing area. The established contributions of the bunches (at $z = 10000$) are thus optimal in phase, but small in absolute value, so the established wakefield is relatively weak.

The picture of bunch contributions after the density step is less clear and even looks chaotic [Fig.\,\ref{f13-spiral}(e,f)]. The general trend is that all contributions are oriented towards the vertical axis, and the main job of the bunches is to push the wave backward. Some of the bunches are accelerated [reside in left half-planes of Fig.\,\ref{f13-spiral}(e,f)], but the decelerated ones are more numerous, so the wave grows, in average. Importantly, bunch contributions never enter the defocusing area and remain strong in absolute value. It is also worth mention that, in the established state [$z = 10000$ in Fig.\,\ref{f13-spiral}(f)], the last bunches are not the most efficient ones, as we noticed when discussing Fig.\,\ref{f10-current}. The general conclusion from this picture is that bunch contributions in the case of the optimized density step are far from being regular and optimum, and there is a room for improvements with more sophisticated density profiles.

\begin{figure*}[tbh]
\includegraphics[width=423bp]{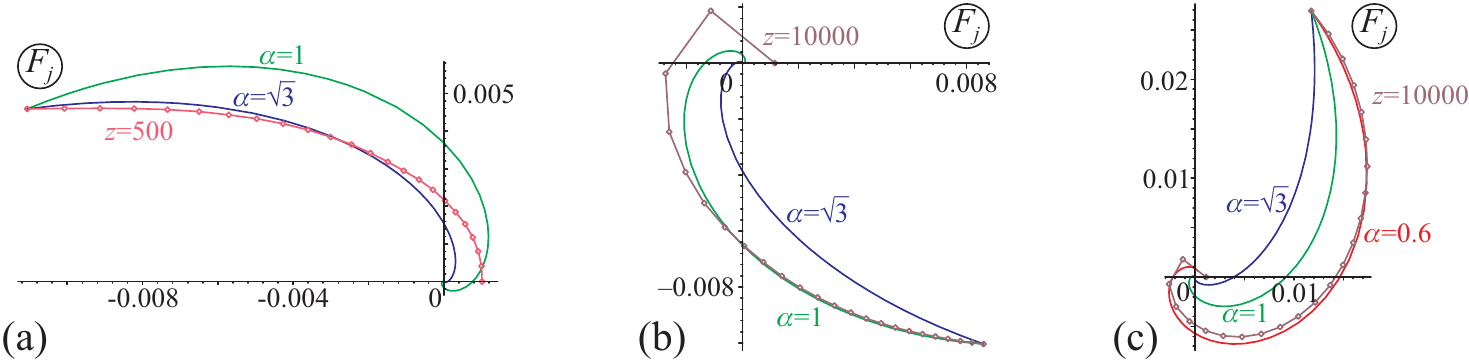}
\caption{The complex amplitude $F_j$ and its theoretical approximations for the uniform plasma at $z=500$ (a) and $z=10000$ (b) and for the optimum density step at $z=10000$ (c).}\label{f14-theory}
\end{figure*}
\begin{figure}[tb]
\includegraphics[width=113bp]{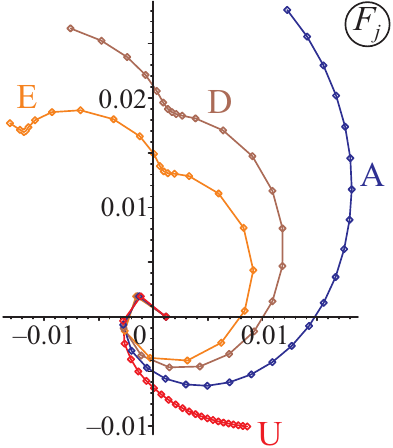}
\caption{The established complex amplitude $F_j$ at $z=10000$ for the uniform plasma and several density steps labeled according to Fig.\,\protect{\ref{f6-steps}}.}\label{f15-other}
\end{figure}

Let us relate the observed bunch contributions with theoretical models. There are two of them available. One model is the wave growth in the linearly responding uniform plasma \cite{PRL107-145003}. It is characterized by the amplitude growth (\ref{e4}) and the phase shift
\begin{equation}\label{e10}
    \delta \phi_j \equiv -\xi_j - 2 \pi j
    = \frac{3 L^{1/3}}{4}  \left( \frac{z}{\tau_b} \right)^{2/3},
\end{equation}
which can be combined to yield the spiral
\begin{equation}\label{e11}
    \Phi_j \propto \exp \left( \alpha \, \delta \phi_j \right), \quad
    F_j \propto \exp \left( (\alpha + i) \delta \phi_j \right)
\end{equation}
with $\alpha = \sqrt{3}$ being the slope of the spiral with respect to a constant radius circle. Another model \cite{AIP396-75} is the optimum train of rectangular bunches that completely fills focusing and decelerating phases of the wave. In the limit of large bunch numbers it forms the spiral (\ref{e11}) with $\alpha = 1$. The theoretical spirals are close to numerically obtained ones at corresponding times (Fig.\,\ref{f14-theory}), though more sensitive diagnostics (maps of $dF_j$) reveals deviations from expected slopes [dashed lines in Fig.\,\ref{f13-spiral}(b,c)]. The established wake in the case of the density step also follows the spiral (\ref{e11}), but with $\alpha \approx 0.6$ [Fig.\,\ref{f14-theory}(c), Fig.\,\ref{f13-spiral}(e)].

At multiples of the optimum step magnitude, some bunches are almost completely destroyed and make no contribution to the wakefield, while others resonantly drive a substantial wake (Fig.\,\ref{f15-other}). This effect is similar to wakefield excitation by a non-resonant bunch train \cite{PPCF52-065009} and involves partial or complete defocusing of the bunches which contribution to the wake is destructive.

\section{Summary} \label{s6}

Let us summarize the key findings of the paper.

In the uniform plasma, the seeded self-modulation instability first transforms the long particle beam into the train of well separated bunches and then partly destroys these bunches. The larger the bunch number, the smaller part of the bunch remains. Both creation and destruction of the bunches occur as a result of particle defocusing by the plasma wakefield. Development of the SMI is accompanied with a flow of beam particles through potential wells of the plasma wave. The wells crawl backward along the beam, grab new portions of beam particles from one side, and release particles from another. As soon as the potential wells run into particle-depleted cross-sections of the beam, the SMI passes into the nonlinear stage, and the wakefield growth slows down. At the propagation distance where the maximum wakefield is reached, the bunches are well separated in space and coherently drive the wave. The wave motion with respect to the beam, however, continues after this point and defocus major parts of the bunches. The remains of the bunches are optimally spaced for stable wakefield excitation, but are too small to excite a strong wave.

The character of beam evolution can be changed by increasing the plasma density at the linear stage of instability growth. The optimum magnitude of the increase is such that to make the beam one wavelength longer if measured in local plasma wavelengths. In the plasma of the increased density, the bunch-to-bunch distance is longer than the plasma wavelength, and the average phase velocity of the wave equals the beam velocity. Nevertheless, well separated bunches are formed, and these bunches quickly come to equilibrium with the wave and excite the strong wakefield over a long propagation distance. The detailed study of the equilibrium bunch train, however, reveals that the wakefield is excited not optimally, so there is a hope that even stronger waves can be excited with more sophisticated plasma density profiles.

\acknowledgments

This work is supported by The Russian Science Foundation, grant No. 14-12-00043. The computer simulations are made at Siberian Supercomputer Center SB RAS.


\begin{thebibliography}{38}
\bibitem{PPCF56-084013}
	R.Assmann, R.Bingham, T.Bohl, C.Bracco, B.Buttenschon, A.Butterworth, A.Caldwell, S.Chattopadhyay, S.Cipiccia, E.Feldbaumer, et al. (AWAKE Collaboration),
	Plasma Phys. Control. Fusion \textbf{56}, 084013 (2014).
 \bibitem{NatPhys9-363}
    A.Caldwell, K.Lotov, A.Pukhov, and F.Simon,
    Nature Phys. {\bf 5}, 363 (2009).
 \bibitem{PRST-AB13-041301}
	K.V.Lotov,
	Phys. Rev. ST Accel. Beams {\bf 13}, 041301 (2010).
 \bibitem{SRep4-4171}
	L.Yi, B.Shen, L.Ji, K.Lotov, A.Sosedkin, X.Zhang, W.Wang, J.Xu, Y.Shi, L.Zhang, and Z.Xu,
	Scientific Reports \textbf{4}, 4171 (2014).
 \bibitem{PRST-AB16-071301}
    L.Yi, B.Shen, K.Lotov, L.Ji, X.Zhang, W.Wang, X.Zhao, Y.Yu, J.Xu, X.Wang, Y.Shi, L.Zhang, T.Xu, Z.Xu,
    Phys. Rev. ST Accel. Beams \textbf{16}, 071301 (2013).
\bibitem{PAC09-4542}
    R. Assmann, M. Giovannozzi, Y. Papaphilippou, F. Zimmermann, A. Caldwell, G. Xia,
    Generation of short proton bunches in the CERN accelerator complex.
    Proceedings of PAC09, (Vancouver, BC, Canada), p.4542--4544.
 \bibitem{PPCF53-014003}
	A. Caldwell, K. Lotov, A. Pukhov and G. Xia,
	Plasma Phys. Controlled Fusion {\bf 53}, 014003 (2011).
 \bibitem{AIP1229-510}
	G. Xia, A. Caldwell, K. Lotov, A. Pukhov, N. Kumar, W. An, W. Lu, W. B. Mori, C. Joshi, C. Huang, P. Muggli, R. Assmann, F. Zimmermann,
	AIP Conf. Proc. \textbf{1229}, 510 (2010).
 \bibitem{IPAC10-4395}
	G.Xia, A. Caldwell,
	Producing Short Proton Bunch For Driving Plasma Wakefield Acceleration.
	Proceedings of IPAC2010 (Kyoto, Japan), p.4395--4397.
 \bibitem{EPAC98-806}
    K.V.Lotov,
    Instability of long driving beams in plasma wakefield accelerators.
    Proc. 6th European Particle Accelerator Conference (Stockholm, 1998), p.806-808.
 \bibitem{PRL104-255003}
	N.Kumar, A.Pukhov, and K.Lotov,
	Phys. Rev. Lett. {\bf 104}, 255003 (2010).
\bibitem{IPAC14-1537}
    R. Tarkeshian,
    Proton electron accelerator at CERN.
	Proceedings of IPAC2014 (Dresden, Germany), p.1537-1539.
\bibitem{NIMA-740-48}
    C.Bracco, E.Gschwendtner, A.Petrenko, H.Timko, T.Argyropoulos, H.Bartosik, T.Bohl, J.E.Mueller, B.Goddard, M.Meddahi, A.Pardons, E.Shaposhnikova, F.M.Velotti, H.Vincke,
    Nucl. Instr. Meth. A \textbf{740}, 48 (2014).
 \bibitem{PoP21-123116}
    K.V.Lotov, A.P.Sosedkin, A.V.Petrenko, L.D.Amorim, J.Vieira, R.A.Fonseca, L.O.Silva, E.Gschwendtner, and P.Muggli,
    Phys. Plasmas \textbf{21}, 123116 (2014).
\bibitem{PRL112-045001}
    Y. Fang, V. E. Yakimenko, M. Babzien, M. Fedurin, K. P. Kusche, R. Malone, J. Vieira, W. B. Mori, and P. Muggli,
    Phys. Rev. Lett. \textbf{112}, 045001 (2014).
\bibitem{NIMA-740-74}
    M.Gross, R.Brinkmann, J.D.Good, F.Gr\"uner, M.Khojoyan, A.Martinez de la Ossa, J.Osterhoff, G.Pathak, C.Schroeder, F.Stephan,
    Nucl. Instr. Meth. A \textbf{740}, 74 (2014).
 \bibitem{AIP1229-467}
	A. V. Petrenko, K. V. Lotov, P. V. Logatchov and A. V. Burdakov,
	AIP Conf. Proc. \textbf{1229}, 467 (2010).
\bibitem{PoP19-063105}
    J. Vieira, Y. Fang, W. B. Mori, L. O. Silva, and P. Muggli,
    Phys. Plasmas \textbf{19}, 063105 (2012).
\bibitem{IPAC14-1476}
    P. Muggli, O. Reimann, L.D. Amorim, N.C. Lopes, J.M. Vieira, S.J. Gessner, M. Hogan, S.Z. Li, M.D. Litos, K. Marsh, W. Mori, C. Joshi, N. Vafaei-Najafabadi, E. Adli, V.K. Berglyd Olsen,
    Electron bunch self-modulation in a long plasma at SLAC FACET.
	Proceedings of IPAC2014 (Dresden, Germany), p.1476-1478.
 \bibitem{NIMA-410-461}
    K.V.Lotov,
    Nuclear Instr. Methods A \textbf{410}, 461 (1998).
 \bibitem{PRL107-145003}
	A. Pukhov, N. Kumar, T. Tuckmantel, A. Upadhyay, K. Lotov, P. Muggli, V. Khudik, C. Siemon, and G. Shvets,
	Phys. Rev. Lett. \textbf{107}, 145003 (2011).
  \bibitem{PRL107-145002}
    C.B.Schroeder, C.Benedetti, E.Esarey, F.J.Gruener, and W. P. Leemans,
    Phys. Rev. Lett. \textbf{107}, 145002 (2011).
 \bibitem{AIP1507-103}
	A.Pukhov, T.Tuckmantel, N.Kumar, A.Upadhyay, K.Lotov, V.Khudik, C.Siemon, G.Shvets, P.Muggli, and A.Caldwell,
	AIP Conf. Proc. \textbf{1507}, 103 (2012).
  \bibitem{PoP19-010703}
    C.B.Schroeder, C.Benedetti, E.Esarey, F.J.Gruener, and W. P. Leemans,
    Phys. Plasmas \textbf{19}, 010703 (2012).
 \bibitem{PoP20-013102}
	K.V.Lotov, A.Pukhov, and A.Caldwell,
	Phys. Plasmas \textbf{20}, 013102 (2013).
 \bibitem{PRST-AB16-041301}
	K.V.Lotov, G.Z.Lotova, V.I.Lotov, A.Upadhyay, T.Tuckmantel, A.Pukhov, A.Caldwell,
	Phys. Rev. ST Accel. Beams \textbf{16}, 041301 (2013).
\bibitem{PoP20-103111}
    C.Siemon, V.Khudik, S.A.Yi, A.Pukhov, and G.Shvets,
    Phys. Plasmas \textbf{20}, 103111 (2013).
\bibitem{Joshi}
    C. Joshi, private communications (2009).
\bibitem{PRE86-026402}
    C.B.Schroeder, C.Benedetti, E.Esarey, F.J.Gruner, and W.P.Leemans,
    Phys. Rev. E \textbf{86}, 026402 (2012).
\bibitem{PoP20-056704}
    C.B.Schroeder, C.Benedetti, E.Esarey, F.J.Gruner, and W.P.Leemans,
    Phys. Plasmas \textbf{20}, 056704 (2013).
\bibitem{PRL112-205001}
    J. Vieira, W. B. Mori, and P. Muggli,
    Phys. Rev. Lett. \textbf{112}, 205001 (2014).
 \bibitem{PoP20-083119}
    K.V.Lotov,
    Phys. Plasmas \textbf{20}, 083119 (2013).
 \bibitem{PoP21-083107}
	K.V.Lotov, V.A.Minakov, and A.P.Sosedkin,
	Phys. Plasmas \textbf{21}, 083107 (2014).
\bibitem{PoP21-056703}
    Y. Fang, J. Vieira, L. D. Amorim, W. Mori, and P. Muggli,
    Phys. Plasmas \textbf{21}, 056703 (2014).
 \bibitem{PRL112-194801}
	K.V.Lotov, A.P.Sosedkin, A.V.Petrenko,
	Phys. Rev. Lett. \textbf{112}, 194801 (2014).
\bibitem{PRL109-145005}
    J.Vieira, R.A.Fonseca, W.B.Mori, and L.O.Silva,
    Phys. Rev. Lett. \textbf{109}, 145005 (2012).
\bibitem{PoP21-056705}
    J. Vieira, R. A. Fonseca, W. B. Mori, and L. O. Silva,
    Phys. Plasmas \textbf{21}, 056705 (2014).
 \bibitem{PoP18-024501}
	K.V.Lotov,
	Phys. Plasmas \textbf{18}, 024501 (2011).
 \bibitem{PoP18-103101}
	A. Caldwell  and K. V. Lotov,
	Phys. Plasmas {\bf 18}, 103101 (2011).
 \bibitem{PoP5-785}
    K.V.Lotov,
    Phys. Plasmas \textbf{5}, 785 (1998).
\bibitem{LCODE}
    See \verb"www.inp.nsk.su/~lotov/lcode" for code description and manual.
\bibitem{PRST-AB6-061301}
    K.V.Lotov,
    Phys. Rev. ST Accel. Beams {\bf 6}, 061301 (2003).
 \bibitem{IPAC13-1238}
	K.V. Lotov, A. Sosedkin, E.Mesyats,
	Simulation of Self-modulating Particle Beams in Plasma Wakefield Accelerators.
	Proceedings of IPAC2013 (Shanghai, China), p.1238-1240.
  \bibitem{PAcc22-81}
    T.Katsouleas, S.Wilks, P.Chen, J.M.Dawson, and J.J.Su,
    Part.Accel. \textbf{22}, 81 (1987).
  \bibitem{PF30-252}
    R.Keinigs and M.E.Jones
     Phys. Fluids \textbf{30}, 252 (1987).
 \bibitem{AIP396-75}
    B.N.Breizman, P.Z.Chebotaev, A.M.Kudryavtsev, K.V.Lotov, and A.N.Skrinsky,
    AIP Conf. Proc. \textbf{396}, 75 (1997).
 \bibitem{PPCF52-065009}
	K.V.Lotov, V.I.Maslov, I.N.Onishchenko, and E.N.Svistun
	Plasma Phys. Control. Fusion \textbf{52}, 065009 (2010).
\end{thebibliography}
\end{document}